\documentclass[11pt]{article}

\long\def\symbolfootnote[#1]#2{\begingroup%
\def\thefootnote{\fnsymbol{footnote}}\footnote[#1]{#2}\endgroup}

 \font\tenrm=cmr10 \font\tenit=cmti10
\font\elevenbf=cmbx10 scaled\magstep 1 \font\elevenrm=cmr10
scaled\magstep 1  1

\newcommand{\x}{\hat{x}}
\newcommand{\be}{\begin{equation}}
\newcommand{\ee}{\end{equation}}
\newcommand{\la}{\label}

\newcommand{\oal}{\overline{\alpha}}
\newcommand{\opi}{\overline{\pi}}
\newcommand{\omu}{\overline{\mu}}
\newcommand{\onu}{\overline{\nu}}
\newcommand{\obe}{\overline{\beta}}
\newcommand{\oga}{\overline{\gamma}}
\newcommand{\ode}{\overline{\delta}}
\newcommand{\ola}{\overline{\lambda}}
\newcommand{\al}{\alpha}
\newcommand{\de}{\delta}

\def\x{x_1}
\def\y{x_2}
\def\xc{\overline{x_1}}
\def\yc{\overline{x_2}}

\def\be{\begin{equation}}
\def\bea{\begin{eqnarray}}

\def\ee{\end{equation}}
\def\eea{\end{eqnarray}}

\def\square{\vcenter{\vbox{\hrule height.5pt \hbox{\vrule
width.5ptheight7pt \kern7pt \vrule width.5pt} \hrule height.5pt}}}

\renewenvironment{thebibliography}[1]
 { \elevenrm
   \begin{list}{\arabic{enumi}.}
    {\usecounter{enumi} \setlength{\parsep}{0pt}
     \setlength{\itemsep}{3pt} \settowidth{\labelwidth}{#1.}
     \sloppy
    }}{\end{list}}
\begin{document}
\begin{center}
\vglue 0.6cm {
 {\elevenbf        \vglue 10pt
    Nonexistence of Petrov type III Space-Times\\
               \vglue 3pt
on which Weyl's Neutrino Equation or Maxwell's Equations\\
\vglue 3pt satisfy Huygens' Principle \symbolfootnote[2]{Published
on Ann. Inst. Henri Poincar\'e (A) Phys. Theorique \textbf{65},
256 (1996)}
\\}
\vglue 1.0cm { R. G. McLenaghan and F. D. Sasse\symbolfootnote[1]{
Present address: Department of Mathematics, CCT-UDESC, 89223-100
Joinville, SC, Brazil, fsasse@joinville.udesc.br}\\}
\baselineskip=13pt {\tenit Department of Applied Mathematics,
University of Waterloo
\\} \baselineskip=12pt {\tenit Waterloo, Ontario, N2L 3G1,
Canada\\}} \vglue 0.8cm

\end{center}
{ \tenrm\baselineskip=11pt
\begin{quote}
{ Abstract.} --- Extending previous results  we  show that there
are no Petrov type III space-times on which either the Weyl
neutrino equation or Maxwell's equations satisfy Huygens'
principle. We prove the result by using Maple's {\tt NPspinor}
package to convert the five-index necessary condition obtained by
Alvarez and W\"unsch
 to  dyad form.
The integrability conditions of the problem lead to a system of
polynomial equations. We then apply Maple's {\tt grobner} package
to show that this system has no admissible solutions.
\end{quote}
 \begin{quote}
 \noindent
{\sc R\'{e}sum\'e.} ---  En prolongeant  des r\'esultats
pr\'ec\'edents, on
 d\'{e}montre qu'il n'existe aucun espace-temps de type III
de Petrov sur lequel l'\'equation de neutrino de Weyl ou les
\'equations de Maxwell satisfait au principe d'Huygens.  Nous
prouvons le r\'esultat par utilisant le logiciel {\tt NPspinor} de
Maple pour transformer la nouvelle condition \`a cinq indices
obtenue par Alvarez et W\"unsch en composants de rep\`ere
spinoriel. A partir des conditions d'int\'egrabilit\'e du
probl\`eme, on obtient un syst\`eme d'\'equations polynomes. Nous
employons donc le logiciel {\tt grobner} de Maple pour d\'emontrer
qu'il n'existe aucune solution admissible de ce syst\`eme.
\end{quote}
}
\section{Introduction}
\rm This paper is the sixth in a series devoted to the solution of
Hadamard's problem for the conformally invariant scalar wave
equation, Weyl's neutrino equation and source-free Maxwell's
equations. These equations can be written respectively as
 \be
 \la{eq1}
   \square\,  u +\frac{1}{6}Ru=0\,,
 \ee
 \be
 \la{eq2}
 \nabla^{A \dot{A}}
\varphi_A=0\,,\ee
\be
 \la{eq3}
 \nabla^{A \dot{A}} \varphi_{A B}=0\,.
 \ee

The conventions and formalism used in this paper are those of
Carminati and McLenaghan \cite{car88} (CM in the sequel). All
considerations in this paper are entirely local. Part of the
calculations presented here were performed using the {\tt
NPspinor} package available in Maple \cite{cza87,cza92}.

Huygens' principle is valid for eqs. (\ref{eq1}), (\ref{eq2}) and
(\ref{eq3}) if and only if for every Cauchy initial value problem
and every point $x_0$ in the 4-dimensional pseudo-Riemannian space
$V_4$, the solution depends only on the Cauchy data in an
arbitrarily small neighborhood of $S \cap C^-(x_0)$, where $S$
denotes the initial surface and $C^-(x_0)$ the past null conoid
from $x_0$ \cite{had23,wun78,gun65b}. Hadamard's problem for
equations (\ref{eq1}), (\ref{eq2}) and (\ref{eq3}), originally
posed only for scalar equations, is that of determining {\it all}
space-times for which Huygens' principle is valid for a particular
equation. As a consequence of the conformal invariance of the
validity of Huygens' principle, the determination may only be
effected up to an arbitrary conformal transformation on the metric
in $V_4$
 \be
  \tilde g_{ab}=e^{2 \varphi} g_{ab}\,,
   \ee
   where
$\varphi$ is an arbitrary real function.

It is known that Huygens' principle is valid for (\ref{eq1}),
(\ref{eq2}) and (\ref{eq3}) on any space-time conformally related
to the exact plane-wave \cite{gun65b,kun68,wun79}, with metric \be
ds^2=2dv \left[ du + [D(v)z^2+\overline{D}(v) \overline{z}^2+
e(v)z \overline{z} ] dv \right] -2 dz d \overline{z}\,, \ee in a
special coordinate system where $D$ and $e$ are arbitrary
functions. These are the only known space-times on which Huygens'
principle is valid for these equations. Furthermore, it has been
shown \cite{gun74,mcl69,wun79} that these are the only conformally
empty space-times on which Huygens' principle is valid.

In the non-conformally empty case several results are known. In
particular for  Petrov type N, the following result has been
proved \cite{car84,car86,alv91b}: {\it Every Petrov
 type N space-time on which
the equations (\ref{eq1}), (\ref{eq2}) and (\ref{eq3}) satisfy
Huygens' principle is conformally related to an exact plane wave
space-time (5)}. For the case of a Petrov type D the following
result has been established \cite{car87,wun89,mcl90}: {\it There
exist no Petrov type D space-times on which equations (\ref{eq1}),
(\ref{eq2}) or (\ref{eq3}) satisfy Huygens' principle}.

In this paper we complete the analysis  for Petrov type III
spacetimes given by CM, in the case of equations (\ref{eq1}) and
(\ref{eq2}). Our main result is
contained in the following theorem:\\

{\bf Theorem 1}. {\it There exist no Petrov type III space-times
on which Weyl's equation (\ref{eq2}) or Maxwell's equations
(\ref{eq3}) satisfy Huygens' principle}.

\section{Previous results}

Let $\Psi_{ABCD}$ denote the Weyl spinor. Petrov type III
space-times are characterized by the existence of a spinor field
$o^A$ satisfying \be \Psi_{ABCD}o^C o^D=0\,,\;\;\;\;
\Psi_{ABCD}o^D \neq 0. \ee Such spinor field is called a {\it
repeated principal spinor} of the Weyl spinor and is determined by
the latter up to an arbitrary variable complex factor. Let $\iota
^A$ be any spinor field satisfying
 \be
  o_A \iota^A=1\,.
\ee The ordered set $\{o_A,\;\iota_A\}$, called a dyad, defines a
basis for the 1-spinor fields on $V_4$. The main results obtained by CM can
then be stated as follows:\\

{\bf Theorem 2}. {\it The validity of Huygens' principle for the
conformally invariant scalar wave equation (\ref{eq1}), or
Maxwell's equations (\ref{eq2}), or Weyl's neutrino equation
(\ref{eq3}) on any Petrov type III space-time implies that the
space-time is conformally related to one in which every repeated
principal spinor field $o_A$ of the Weyl spinor is recurrent, that
is} \be o_{A;B \dot{B}}=o_AI_{B \dot{B}}\,, \ee where $ I_{B
\dot{B}} $ is a 2-spinor, and \be \Psi_{ABCD;E \dot{E}}\,\iota^A
\iota^B \iota^C o^D o^E \overline{o}^{\dot{E} }=0\,,
 \ee
  \be
R=0\,,\;\;\;\;\;\;\Phi_{AB \dot{A} \dot{B}}o^A o^B=0\,.
 \ee

{\bf Theorem 3.} {\it If any one of the following three
conditions}
 \be
 \la{11}
  \Psi_{ABCD;E \dot{E}}\, \iota^A \iota^B \iota^D
\iota^E \overline{o}^ {\dot{E}}=0\,, \ee \be \la{12} \Psi_{ABCD;E
\dot{E}}\, \iota^A \iota^B o^D o^E \overline{\iota}^
{\dot{E}}=0\,, \ee \be \la{13} \Psi_{ABCD;E \dot{E}}\, \iota^A
\iota^B \iota ^D o^E \overline{o}^ {\dot{E}}=0\,,
\ee
 {\it is
satisfied, then there exist no Petrov type III space-times on
which the conformally invariant scalar wave equation (\ref{eq1})
 or Maxwell's equations (\ref{eq2}), or
Weyl's neutrino equation (\ref{eq3}) satisfies Huygens'
principle.}

It will be proved here that conditions (\ref{11}) to (\ref{13})
are superfluous in  the cases of equations (\ref{eq2}) and
(\ref{eq3}), i.e., they are consequences of the necessary
conditions for the validity of Huygens' principle, in particular
of the five-index necessary conditions derived by Alvarez and
W\"{u}nsch \cite{alv91a,alv91b}.

\section{Necessary conditions}

In order to prove the Theorem 1 we shall need the following
necessary conditions for the validity of Huygens' principle
\cite{gun52,wun70,mcl74,mcl72,wun78}:
 \be
  \la{III}
(III)\;\;\;\;\;\;\;\;\;\;\;S_{abk;}{}^{k}\,-\,
\frac{1}{2}C^{k}{}_{ab}{}^{l}L_{kl}\,=\,0\,,
 \ee

$$
(V)\;\;\;\;\;\; \; TS \left[
k_1C^k{}_{ab}{}^l{}_{;}{}^mC_{kcdl;m}\,+\,2k_2C^k{}_{ab}{}^l{}_{;c}
S_{kld}+2(8k_1-k_2)S_{ab}{}^k{}S_{cdk} \right.
$$
$$
-2k_2C^k{}_{ab}{}^lS_{klc;d}-8k_1C^k{}_{ab}{}^lS_{cdk;l}\,+\,
k_2C^k{}_{ab}{}^lC_l{}^m{}_{ck}L_{dm}
$$
\be \la{V} +\left. 4k_1C^k{}_{ab}{}^lC^m{}_{cdl}L_{km} \right]
\,=\,0\,, \ee where \be C_{abcd}:=R_{abcd}-2g_{[a[d}L_{b]c]}\,,
\ee \be L_{ab}:=-R_{ab}+\frac{1}{6}R g_{ab}\,, \ee \be
S_{abc}:=L_{a[b;c]}\,. \ee
 In the above $C_{abcd}$ denotes the
Weyl tensor, $R_{ab}$ the Ricci tensor and $TS[\;\;]$ the operator
which takes the trace free symmetric part of the enclosed tensor.
In (\ref{V}) $k_1$ takes values 3, 8, 5 and $k_2$
 the values 4, 13, 16, depending on whether the equation under consideration
is the conformally invariant scalar equation, Weyl's equation or
Maxwell's equations respectively.

We shall also need a necessary condition involving five free
indices, valid for Weyl's equation (2) and Maxwell's equations
(3), that was found by Alvarez and W\"{u}nsch
\cite{alv91a,alv91b}. It can be written in the form
 \be
  \la{19}
 T^{(1)}{}_{abcde}+\sigma_1 T^{(2)}{}_{abcde} + \sigma_2
 T^{(3)}{}_{abcde} =0\,, \ee where $\sigma_1 $ and $\sigma_2$ are
 fixed real numbers, and
$$
T^{(1)}{}_{abcde}=TS [4{}^{\ast}C^k{}_{ab}{}^lC^u{}_{del;ukc} -6
^*C^k{}_{bc}{}^l{}_{;a}\, C^u{}_{del;uk} +26
^*C^u{}_{ab}{}^k{}_{;u} \,C^v{}_{dek;vc}+
$$
$$
^*C^k{}_{ab}{}^l{}_{;}{}^n\,C_{k del;nc} +5
^*C^k{}_{ab}{}^l{}\,C^n{}_{del;c}{}L_{kn}+4 ^*C^k{}_{ab}{}^l
C^n{}_{cdl;k}L_{en}
$$
\be \la{20} +4
^*C^k{}_{ab}{}^l\,C^u{}_{klc;u}\,L_{de}-21^*C^k{}_{ab}{}^l\,
C^u{}_{cdk;u}\,L_{el} ]\,, \ee \be \la{21} T^{(2)}{}_{abcde}:=TS
[-12 ^*C^k{}_{ab}{}^l\,C_{kcd}{}^n \, C^u{}_{lne;u}-
C_k{}^{nh}{}_a C_{lnhb}{}^*C^k{}_{de}{}^l{}_{;c}]\,, \ee \be
\la{22} T^{(3)}{}_{abcde}:=TS [-8 ^*C^k{}_{ab}{}^lC_{kcd}{}^n
C^u{}_{lne;u}+ ^*C_k{}^{nh}{}_a
C_{lbnh}\,C^k{}_{de}{}^l{}_{;c}]\,, \ee where \be \la{23}
^*C_{abcd}:=\frac{1}{2}\epsilon_{ab}{}^{ef}C_{efcd}\,. \ee
 It is
worth to mention that, in our conventions, the Riemann tensor,
Ricci tensor and the Ricci scalar have opposite sign to those used
by Alvarez and W\"{u}nsch \cite{alv91b}. The spinor equivalents of
conditions III and V are given, respectively, by
\cite{mcl90,car86}
 \be
 \la{eq-III-V}
\Psi_{ABKL;}{}^K{}_{\dot{A}}{}^L{}_{\dot{B}}+
\Psi_{AB}{}^{KL}\,\Phi_{KL \dot{A} \dot{B}}=0\,, \ee
\begin{eqnarray}
\nonumber
 & & \mbox{}
k_1 \Psi_{ABCD;K \dot{K}} \overline{\Psi}_{\dot{A} \dot{B} \dot{C}
\dot{D};} {}^{\dot{K}K}+k_2
\Psi^K{}_{(ABC;D)(\dot{A}}\overline{\Psi}_{\dot{B} \dot{C}
\dot{D}) \dot{L};}{}^{\dot{L}}{}_K
\\
\nonumber
 & & \mbox{}  +k_2 \overline{\Psi}^{\dot{K}}{}_{(\dot{A} \dot{B} \dot{C};\dot{D})(A}\Psi_{BCD)L;}{}
^L{}_{\dot{K}}- 2(8k_1-k_2)\Psi_{ABC|K|;}{}^K{}_{(\dot{A}}
\overline{\Psi}_{\dot{B} \dot{C} \dot{D})
\dot{K};}{}^{\dot{K}}{}_{D)}
\\
\nonumber
 & & \mbox{} -k_2 \Psi^K{}_{(ABC}\, \overline{\Psi}_{(\dot{A} \dot{B} \dot{C} |\dot{L}|;}
{}^{\dot{L}}{}_{|K| D) \dot{D})} -k_2
\overline{\Psi}^{\dot{K}}{}_{(\dot{A} \dot{B} \dot{C} }
\Psi_{(ABC|L|;}{}^L{}_{|\dot{K}|D) \dot{D})}
\\
\nonumber
 & & \mbox{} +4k_1 \Psi^K{}_{(ABC} \overline{\Psi}_{\dot{A} \dot{B} \dot{C} |\dot{L}|;}
{}^{\dot{L}}{}_{D)K \dot{D})} +4k_1
\overline{\Psi}^{\dot{K}}{}_{\dot{A} \dot{B} \dot{C}} \Psi_{
ABC|L|;}{}^L{}_{\dot{D})D)\dot{K}}
\\
 & & \mbox{} +2(k_2-4k_1) \Psi^K{}_{(ABC} \Phi_{D)K \dot{K} (\dot{A}} \overline{\Psi}_
{\dot{B} \dot{C} \dot{D})}{}^{\dot{K}} -2(4k_1+k_2) \Lambda
\Psi_{ABCD} \overline{\Psi}_{\dot{A} \dot{B} \dot{C} \dot{D}}=0\,.
\end{eqnarray}

\section{Proof of Theorem 1}

In {\bf CM}, Theorem 2 was proved by using  conditions III and V.
The explicit form of these necessary conditions is obtained by
first converting the spinorial expressions to the dyad form and
then contracting them with appropriate products of $o^A$ and
$\iota ^A$ and their complex conjugates. In particular, it was
shown, that there exists a dyad $\{o_A, \iota_A\}$
 and a conformal transformation
such that
\be
 \la{26}
  \kappa=\sigma=\rho=\tau=\epsilon=0\,, \ee
\be
 \la{27}
  \Psi_0=\Psi_1=\Psi_2=\Psi_4=0, \;\;\;\; \Psi_3=-1\,,
\ee
 \be
  \la{28}
  \Phi_{00}=\Phi_{01}=\Phi_{02}=R=0\,,
   \ee
    \be
\la{29}
 D\alpha=D\beta=0\,,
  \ee
  \be
   \la{30}
    \Phi_{11}=c\,,
     \ee
where $c$ is a constant. By contracting condition (III) with
$\iota^{AB} \overline{o}^{\dot{A} \dot{B}}$ and $\iota^A o^B
\overline{\iota}^{\dot{A} \dot{B}}$ (where the notation \linebreak
$o_{A_1 \cdots A_p}=o_{A_1} \cdots o_{A_p}$, etc. has been used)
we get, respectively,
 \be
 \la{31}
  D\pi=0\,, \ee \be \la{32}
\delta\beta=-\beta(\overline{\alpha} + \beta) \,.
\ee
 From the Bianchi
identities, using the above conditions, we obtain \be \la{33}
D\Phi_{12}=2 \overline{\pi} \Phi_{11}\,, \ee \be \la{34}
D\Phi_{22}=-2(\beta+\overline{\beta})+2\Phi_{21} \overline{\pi}+
2\Phi_{12} \pi\,, \ee
 \be
  \la{35}
\delta\Phi_{12}=2\overline{\alpha}+4\overline{\pi}+2\overline{\lambda}
\Phi_{11} -2 \overline{\alpha}\Phi_{12}\,, \ee \be \la{36}
\overline{\delta}\Phi_{12}=-2\beta+2\overline{\mu} \Phi_{11}-
2\overline{\beta} \Phi_{12}\,,
 \ee
$$
\de \Phi_{22}=\Delta\Phi_{12}+2 \oga +4\omu -2\onu \Phi_{11}+2\ola
\Phi_{21}
$$
\be
 \la{36a}
 +2\Phi_{12} \mu -2 \Phi_{22} \beta-2\Phi_{22} \oal
+2\oga \Phi_{12}\,.
 \ee
  Using Ricci identities we get
 \be \la{37a} D\gamma=\opi \al
+\beta \pi+\Phi_{11}\,, \ee \be \la{37b}
\delta\opi=D\ola-\opi^2-\opi \oal+\opi\beta\,, \ee \be \la{37c}
D\onu=\Delta\opi+\opi \omu +\ola
\pi+\opi\oga-\opi\gamma-1+\Phi_{12}\,, \ee \be \la{37}
\delta\alpha=\ode \beta+\alpha \overline{\alpha}+ \beta
\overline{\beta}-2\beta \alpha + \Phi_{11}\,, \ee
\be
\la{40}
 \delta\pi=D\mu-\opi \pi +\pi \oal-\beta \pi\,.
  \ee
We can obtain useful integrability conditions for the above
identities by using NP commutation relations. Using (\ref{33}),
(\ref{34}), (\ref{36a}), (\ref{32}), (\ref{37a}), (\ref{37b}),
(\ref{37c}) and (\ref{40}) in the commutator expression
$[\delta,D] \Phi_{22}-[\Delta,D] \Phi_{12}$, gives \be \la{43}
\delta\obe=-2\Phi_{11}-\obe \oal - 4\obe \opi-2D\omu-\beta \obe +2
\opi \pi. \ee

We shall consider both Maxwell and Weyl cases separately. We begin
with the Maxwell equations, i.e., $k_1=5$ and $k_2=16$. By
contracting condition V with $ \iota^{ABCD}
\overline{\iota}^{\dot{A} \dot{B}} \overline{o}^{\dot{C}
\dot{D}}$, we get \be \la{44} 62\obe
\pi+40\obe\al+6\pi\al+3\al^2+\ode\al+2\ode\pi+\ode\obe+\obe^2=0\,.
\ee
 Substituting (\ref{32}) into this equation results in
  \be
   \la{45}
 \de(2\opi+\oal)=-62\opi\beta-39\beta\oal-6\opi\oal-3\oal^2\,.
  \ee
  Now, from  (\ref{37}), (\ref{40}) and (\ref{43}) we obtain
  \be
   \la{46}
 \delta(2\pi +\alpha)=2\pi \oal +\alpha \oal -6\beta \pi
 -3\beta\alpha-\Phi_{11}\,.
  \ee
Contracting  condition V with $ \iota^{ABC} o^D \overline{\iota}^
{\dot{A} \dot{B} \dot{C} } \overline{o}^{\dot{D}}$, using
(\ref{37a}), (\ref{40}) and the complex conjugate of (\ref{43}),
we get \be \la{47} 148 \Phi_{11}+152 \obe \opi + 76(D\mu+D\omu) -8
\opi \al -104 \beta \obe -8 \al \oal -232 \opi \pi +152 \beta \pi
-8 \pi \oal =0\,. \ee Using (\ref{45}), (\ref{46}), (\ref{32}),
(\ref{37}), (\ref{40}), (\ref{43}) and (\ref{47}) in
$[\ode,\delta](\alpha+2\pi)=(\alpha-\obe)
\delta(\alpha+2\pi)+(-\oal + \beta)\ode(\alpha+2 \pi) $, we obtain
one expression for $D\omu$. Substituting it in (\ref{47}) and
solving for $D\mu$ we obtain
\begin{eqnarray}
\la{48} \nonumber \lefteqn{{D\mu}=\frac{1}{152} \left(-1520 \al
\beta \pi +208 \al \oal \pi
-152 \obe \opi \pi +1040 \beta \obe \al + 1968 \obe \beta \pi \right.}\\
\nonumber
 & &  \mbox{} +1216 \obe \pi \oal -1228 \pi \Phi_{11} +1688 \al \pi \opi
 -739 \al \Phi_{11} +80 \al^2 \oal +380 \obe \Phi_{11} \\
 & &  \mbox{} \left. +2496 \opi \pi^2 -2432 \beta \pi^2 +128 \pi^2 \oal +80 \opi \al^2
 +760 \obe \al \oal
\right)/(-\obe+5\al+8\pi)\,,
\end{eqnarray}
where we have assumed that $\obe -5 \al -8 \pi \not = 0$. By
substituting this equation into (\ref{46}), we find:
\begin{eqnarray}
\la{49} \nonumber \lefteqn{{D\omu}=\frac{1}{152} \left( 776 \al
\obe \oal -304 \obe^2 \opi +208 \obe^2 \beta +84 \obe \Phi_{11}
-1216 \opi \pi^2
 \right. }
\\
\nonumber
 & &  \mbox{}
+741 \al \Phi_{11} + 1536 \al \opi \obe-760 \al \opi \pi  + 2744
\opi \pi \obe +1332 \pi \obe \oal
\\
 & &  \mbox{}
\left. +741 \al \Phi_{11} +1140 \pi \Phi_{11}
\right)/(\obe-5\al-8\pi)\,.
\end{eqnarray}
We note that (\ref{48}) and (\ref{49}) have the same denominator.
So, in what follows we shall use the Pfaffians $\delta \al$,
$\delta \pi$, and $\delta \obe$, given by (\ref{37}), (\ref{40})
and (\ref{43}), respectively, and their complex conjugates, in
such a way that they have all the same denominator. This procedure
simplifies the expressions to be obtained from the integrability
conditions.

Now need to convert (\ref{19}) to the spinor form in the dyad
basis. The resulting expression, obtained using Maple's package
{\tt NPspinor} \cite{cza92}, has a considerable size, specially
due to the term  in (\ref{20}) containing a third order derivative
of the Weyl tensor, and will not be presented here. However, among
the twenty one independent spinor components, we found a
relatively simple one, obtained by contracting the dyad expression
with $\iota^{AB} o^{CDE} \overline{\iota}^{\dot{A} \dot{B} \dot{C}
\dot{D} \dot{E}}$, It has the following form:
\begin{eqnarray}
\la{50} \nonumber & &  \mbox{} -14\oal\
\de\opi-12\oal^3-\de(\de(\oal+2\opi))-21 \beta\oal^2
+7\ola\beta^2+14\beta^2\opi
\\
 & &  \mbox{} -7\beta\de(\oal+2\opi)-42\beta\oal\opi
-(10\oal+6\opi)\de\oal-24\oal^2\opi=0 \, .
\end{eqnarray}
We observe that the terms (\ref{21}) and (\ref{22}) did not
contribute to this component. Using (\ref{45}) to eliminate $\de
\pi$ from this equation, and solving for $\de\oal$ we get
 \be
 \la{51}
  \delta\oal=192 \beta \opi -3 \oal^2 +121 \beta \oal\,
  \ee
and
 \be
  \la{52}
   \de\opi=-127 \opi \beta -80\beta \oal -3 \opi \oal\,.
 \ee
We have now all the Pfaffians we need to complete the proof. The
integrability conditions provided by the NP commutation relations
can now be used. Let us consider  the NP commutator
$[\ode,\delta]\al $. Using the Pfaffians calculated previously,
and
 solving for $\Phi_{11}$, we obtain
\small
\begin{eqnarray}
\la{53} \nonumber \lefteqn{{\Phi_{11}} :=  - 8\obe\,(\,8\,\pi +
5\,\al\,)(172736 \,\opi\,\pi^{2} - 7776\,\pi^{2}\,\oal -
13294\,\al\,\pi\,
\oal + 7866\,\obe\,\pi\,\oal} \\
\nonumber
 & & \mbox{} + 211556\,\al\,\opi\,\pi + 9568\,\obe\,\beta\,
\pi - 22572\,\obe\,\opi\,\pi + 5330\,\obe\,\beta\,\al
 + 4845\,\obe\,\al\,\oal \\
\nonumber
 & & \mbox{} - 1805\,\obe^{2}\,\oal - 2470\,\beta\,\obe
^{2} - 13110\,\opi\,\al\,\obe - 5290\,\al^{2}\,\oal
 + 65010\,\opi\,\al^{2})/  \\
 & & \mbox{}  (-772320\,\pi^{2}\,\obe - 2048352\,\al\,\pi^{2} -
1158240\,\pi\,\al^{2} + 4085\,\al\,\obe^{2} + 10868\,\pi\,
\obe^{2} \\
 & & \mbox{}
- 335985\,\al^{2}\,\obe - 1020276\,\al\,\pi\, \obe +
214700\,\al^{3} + 1191680\,\pi^{3})\,.
\end{eqnarray}
\normalsize On the other hand, from the  commutator
$[\ode,\de](\al+\obe)$ the following expression for $\Phi_{11}$
results: \small
\begin{eqnarray}
\la{54} \nonumber \lefteqn{{\Phi_{11}} := - 8\obe( -
20672\,\obe\,\opi\,
\pi - 12920\,\opi\,\al\,\obe + 8056\,\obe\,\pi\,{\oal} + 172736\,\opi\,\pi^{2}} \\
\nonumber
 & & \mbox{} - 7776\,\pi^{2}\,\oal + 9568\,\obe\,\beta\,\pi
 + 5035\,\obe\,\al\,\oal - 5290\,\al^{2}\,\oal +
65010\,\opi\,\al^{2} \\
 & & \mbox{} - 13294\,\al\,\pi\,\oal + 211556\,\al\,\opi
\,\pi + 5330\,\obe\,\beta\,\al) /(- 13047\,\obe\,\al\\
 & & \mbox{}
   + 42940\,\al^{2} + 162944\,\pi\,\al
 + 10412\,\obe^{2} - 18564\,\pi\,\obe + 148960\,\pi^{2}
)\,.
\end{eqnarray}
\normalsize Using the fact that $\ode(\Phi_{11})=0$, we obtain,
from (\ref{54}), a third expression for $\Phi_{11}$:
\small
\begin{eqnarray}
\la{55} \lefteqn{{\Phi_{11}} := 8\obe( -
48191081692160\,\pi^{3}\,{\oal
}\,\al^{2} + 180931104170496\,\opi\,\pi^{3}\,\al^{2}} \\
 & & \mbox{} + 7968511840\,\obe^{4}\,\pi\,\oal -
265004094784\,\obe^{3}\,\beta\,\pi^{2} - 5268590200832\,{\opi
}\,\pi^{4}\,\obe \\ \nonumber
 & & \mbox{} + 24894675520\,\obe^{4}\,\beta\,\pi + 221453789400
\,\opi\,\al^{2}\,\obe^{3} - 20739582848\,\obe^{3}\,{p }^{2}\,\oal
\\ \nonumber
 & & \mbox{} + 4980319900\,\obe^{4}\,\al\,\oal -
30330200378072\,\pi^{2}\,\oal\,\al^{3} - 1208861900450\,\al^{
5}\,\oal \\ \nonumber
 & & \mbox{} + 5225292181050\,\opi\,\al^{5} + 724772046800\,
\obe^{3}\,\opi\,\pi\,\al \\ \nonumber
 & & \mbox{} - 3675701240760\,\obe^{2}\,\opi\,\pi\,\al^{2
} - 6145239989312\,\obe^{2}\,\opi\,\pi^{2}\,\al \\ \nonumber
 & & \mbox{} - 7649757648780\,\obe\,\opi\,\pi\,\al^{3} -
16323677160464\,\obe\,\opi\,\pi^{2}\,\al^{2} \\ \nonumber
 & & \mbox{} - 15274850502912\,\obe\,\opi\,\pi^{3}\,\al
 - 26601021440\,\obe^{3}\,\pi\,\oal\,\al \\ \nonumber
 & & \mbox{} - 1040399202440\,\obe^{2}\,\pi\,\oal\,\al^{2
} - 1635701635136\,\obe^{2}\,\pi^{2}\,\oal\,\al \\ \nonumber
 & & \mbox{} + 5632482563850\,\obe\,\pi\,\oal\,\al^{3} +
13162176133400\,\obe\,\pi^{2}\,\oal\,\al^{2} \\ \nonumber
 & & \mbox{} + 13687783600768\,\obe\,\pi^{3}\,\oal\,\al
 - 329130147840\,\obe^{3}\,\beta\,\pi\,\al \\ \nonumber
 & & \mbox{} + 265058659320\,\obe^{2}\,\beta\,\pi\,\al^{2} +
469041329536\,\obe^{2}\,\beta\,\pi^{2}\,\al \\ \nonumber
 & & \mbox{} + 1952913512680\,\obe\,\beta\,\pi\,\al^{3} +
4375379783424\,\obe\,\beta\,\pi^{2}\,\al^{2} \\ \nonumber
 & & \mbox{} + 4338917366784\,\obe\,\beta\,\pi^{3}\,\al -
102469061500\,\obe^{3}\,\beta\,\al^{2} + 49609754350\,\obe
^{2}\,\beta\,\al^{3} \\ \nonumber
 & & \mbox{} + 325649974650\,\obe\,\beta\,\al^{4} + 15559172200
\,\obe^{4}\,\beta\,\al - 38360907652096\,\pi^{4}\,\oal\,\al
 \\ \nonumber
 & & \mbox{} + 39628423187260\,\opi\,\al^{4}\,\pi -
1329967209650\,\opi\,\al^{4}\,\obe - 858758431040\,{\obe}^{2}\,\pi^{3}\,\oal \\
\nonumber
 & & \mbox{} - 20447502080\,\obe^{4}\,\opi\,\pi -
8524238850\,\obe^{3}\,\al^{2}\,\oal - 12779688800\,{\opi}\,\al\,\obe^{4} \\
 \nonumber
 & & \mbox{} + 905003032125\,\obe\,\al^{4}\,\oal +
136081885849600\,\opi\,\pi^{4}\,\al \\ \nonumber
 & & \mbox{} + 119915073751888\,\opi\,\pi^{2}\,\al^{3} -
9564050953390\,\al^{4}\,\oal\,\pi \\ \nonumber
 & & \mbox{} + 5344911334400\,\pi^{4}\,\oal\,\obe -
12237656133632\,\pi^{5}\,\oal + 1606267826176\,\obe\,\beta
\,\pi^{4} \\ \nonumber
 & & \mbox{} + 40801870077952\,\opi\,\pi^{5} - 730927683900\,
\opi\,\al^{3}\,\obe^{2} - 220969445675\,\obe^{2}\,\al ^{3}\,\oal
\\ \nonumber
 & & \mbox{} + 274949282816\,\obe^{2}\,\beta\,\pi^{3} +
592713574016\,\obe^{3}\,\opi\,\pi^{2} - 3416931669632\,
\obe^{2}\,\opi\,\pi^{3})  \left/ {\vrule height0.37em width0em
depth0.37em} \right.\\ \nonumber
 & & \mbox{} [(\, - 13047\,\obe\,\al + 42940\,\al^{2} + 162944\,\pi\,{
a} + 10412\,\obe^{2} - 18564\,\pi\,\obe + 148960\,\pi^{2} \,)( \\
\nonumber
 & & 102596352\,\pi^{3} - 7684576\,\pi^{2}\,\obe + 194465152
\,\al\,\pi^{2} + 24320\,\pi\,\obe^{2} - 9460852\,\al\,\pi\,\obe
\\
 & & \mbox{}
  \mbox{} + 123050612\,\pi\,\al^{2} + 25995895\,\al^{3} -
133000\,\al\,\obe^{2} - 2915745\,\al^{2}\,\obe)]\,.
\end{eqnarray}

\normalsize We now  suppose that the  denominators in these three
expressions for $\Phi_{11}$ are different from zero. Later we
consider the cases in which each of them is different from zero.
We also suppose that spin coefficients $\al$, $\beta$, $\pi$ are
different from zero. It was shown in CM that if one of these spin
coefficients is equal to zero, the others must be zero too, and
Huygens' principle is violated.

The next step consists in proving that (\ref{53}), (\ref{54}), and
(\ref{55}) imply that $\al$, $\beta$ and $\pi$ are proportional to
each other. In order to get a system of  with only two complex
variables, instead of three, new variables are defined as follows:
\be \la{x_1} x_1:=\frac{\al}{\pi}\,,\quad
x_2:=\frac{\beta}{\opi}\,. \ee
 By subtracting (\ref{53}) from
(\ref{54}), taking the numerator and dividing by
$(8-\y-5\x)(5776\yc^2\pi\opi)$, we get \small
\begin{eqnarray}
\nonumber \lefteqn{{N_1} := 178100\,\x\,\y\,\yc^{2} +
284960\,\y\,\yc^{2} + 208240\,\yc^{2}\,{\xc
} + 130150\,\xc\,\yc^{2}\,\x} \\
\nonumber
 & & \mbox{} + 109825\,\xc\,\yc\,\x^{2} +
252850\,\y\,\yc\,\x^{2} + 523744\,\y\,
\yc + 3451480\,\x\,\yc \\
\nonumber
 & & \mbox{} + 341900\,\yc\,\x\,\xc + 265888\,
\yc\,\xc + 735800\,\y\,\yc\,\x +
2915264\,\yc \\
\nonumber
 & & \mbox{} + 1018400\,\yc\,\x^{2} - 879008\,{\xc} +
18263488 - 408050\,\x^{3}\,\xc + 4864450\,
\x^{3} \\
 & & \mbox{} + 35335248\,\x + 22731900\,\x^{2} -
2101032\,\x\,\xc - 1622550\,\x^{2}\,\xc\,=0 \,.
\end{eqnarray}
\normalsize
 Subtracting (\ref{54}) from (\ref{55}), taking the
numerator and dividing by $(8-\y-5\x)(152\pi\opi)$, gives \small
\begin{eqnarray}
\nonumber \lefteqn{{N_2} :=  - 11651821200\,\x^{3} - 26531539120\,
\x^{2} - 10132263424 - 26800626944\,\x} \\
\nonumber
 & & \mbox{} + 242619584\,\yc + 593671488\,\yc^{2} +
2256829184\,\xc - 35155250\,\yc^{2}\,\y\,{\x}^{2} \\
\nonumber
 & & \mbox{} + 21550100\,\yc^{3}\,\y\,\x -
128589500\,\yc\,\y\,\x^{3} + 11036720\,{\yc
}^{3}\,\xc \\
\nonumber
 & & \mbox{} + 21755825\,\yc^{2}\,\x^{2}\,\xc
 + 130839250\,\x^{3}\,\yc - 216631750\,\yc\,
\x^{3}\,\xc \\
\nonumber
 & & \mbox{} - 17700400\,\yc^{3}\,\x + 372958500\,
\x^{4}\,\xc + 220600700\,\yc^{2}\,\x^{2}
 + 34480160\,\y\,\yc^{3} \\
\nonumber
 & & \mbox{} - 1915591500\,\x^{4} - 28320640\,\yc^{3}
 + 6897950\,\yc^{3}\,\x\,\xc + 724005800\,{\yc}^{2}\,\x \\
\nonumber
 & & \mbox{} - 112640320\,\x\,\y\,\yc^{2} -
90878112\,\y\,\yc^{2} + 59839936\,\yc^{2}\,
\xc \\
\nonumber
 & & \mbox{} + 72209280\,\xc\,\yc^{2}\,\x -
1002142990\,\xc\,\yc\,\x^{2} - 585808600\,{\y}\,\yc\,\x^{2} \\
 & & \mbox{} - 448045312\,\y\,\yc + 632690016\,{\x
}\,\yc - 1546809184\,\yc\,\x\,\xc -
796702240\,\yc\,\xc \\
\nonumber
 & & \mbox{} - 887976960\,\y\,\yc\,\x +
509836460\,\yc\,\x^{2} + 2328634300\,\x^{3}\,
\xc \\
\la{58}
 & & \mbox{} + 5728848896\,\x\,\xc + 5470002280\,
\x^{2}\,\xc\,=0\,.
\end{eqnarray}
\normalsize

At this point we shall consider $\x$, $\y$, $\xc$, $\yc$ as
independent variables, and use the package {\tt grobner} in Maple,
for the polynomial system formed by the polynomials $N_1$, $N_2$.
This package computes a collection of reduced (lexicographic)
Gr\"obner bases corresponding to  a set of polynomials. The result
is a list of reduced subsystems whose roots are those of the
original system, but whose variables have been successively
eliminated and
  separated as far as possible. In the present
case we obtain four subsystems given by \small
\[
{{G}_{1}} := [\,42 - 3\,\xc + 65\,\y\,\yc, 8 + 5\,\x\,]\,,
\]
\begin{eqnarray*}
\lefteqn{{G_2} := [205049562510\,\xc\,\y -
2072817918600\,\y\,\yc + 529175067720\,\y} \\
 & & \mbox{} + 5001500073283\,\xc^{2} - 3239213905470\,
\xc + 3029503111800\,\yc \\
 & & \mbox{} - 26163100475032, 23707187714600\,\y\,{\yc
}^{2} - 12070111345240\,\y\,\yc \\
 & & \mbox{} - 15004500219849\,\xc - 34648966659800\,{\yc}^{2} + 11975391986580\,\yc \\
 & & \mbox{} + 41386627076564,  - 1113092 - 431311\,\xc +
1909780\,\yc + 954890\,\yc\,\xc,  \\
 & & 205\,\x + 368]\,,
\end{eqnarray*}

\begin{eqnarray*}
\lefteqn{{{G}_{3}} := [2175607695654600868570\,\xc\,{\y
} - 244429060944194171242925\,\y\,\yc\,\x} \\
 & & \mbox{} - 362016456337543432617920\,\y\,\yc -
25492004395136420363950\,\y\,\x \\
 & & \mbox{} - 33777552239002428460240\,\y -
352210319977170626297190\,\x\,\xc^{2} \\
 & & \mbox{} - 527515033185400238012371\,\xc^{2} -
372609773697867989940085\,\x\,\xc \\
 & & \mbox{} - 568758266358009596992694\,\xc +
357242473687668404124275\,\x\,\yc \\
 & & \mbox{} + 529100974647178863056960\,\yc +
421165196163010815629650\,\x \\
 & & \mbox{} + 613523694569903050334320, 74421671368200\,\y
\,\yc^{2} \\
 & & \mbox{} - 372108356841000\,\y\,\yc\,\x -
595373370945600\,\y\,\yc \\
 & & \mbox{} + 202992871981785\,\x\,\xc +
309188233840256\,\xc - 108770135076600\,\yc^{2} \\
 & & \mbox{} + 408707348737250\,\x\,\yc +
731690446259960\,\yc - 363026773505180\,\x \\
 & & \mbox{} - 558465136160528,  \\
 & & 139740\,\yc\,\xc - 497365\,\x\,\xc
 - 799324\,\xc + 279480\,\yc - 1187280\,\x -
1879248, \\
 & &  43975\,\x^{2} + 137900\,\x + 107824]\,,
\end{eqnarray*}

\begin{eqnarray*}
\lefteqn{{{G}_{4}} := [15138500\,\xc\,\y\,\yc
 - 200682625\,\y\,\x^{2}\,\xc - 589775940\,
\y\,\x\,\xc} \\
 & & \mbox{} - 425769864\,\xc\,\y + 30277000\,{\y
}\,\yc - 677199250\,\y\,\x^{2} - 2061417280\,
\y\,\x \\
 & & \mbox{} - 1553711328\,\y - 34073270\,\yc\,{\xc}^{2}
- 2202762525\,\x^{2}\,\xc^{2} - 6761833290\,
\x\,\xc^{2} \\
 & & \mbox{} - 5193018500\,\xc^{2} - 13933181776 -
6178746050\,\x^{2} - 18555219840\,\x \\
 & & \mbox{} - 180544080\,\yc - 17865753288\,\xc -
158418580\,\yc\,\xc - 23445716600\,\x\,{\xc} \\
 & & \mbox{} - 7696469075\,\x^{2}\,\xc, 38482345375\,
\x^{3}\,\xc + 30893730250\,\x^{3} \\
 & & \mbox{} + 11013812625\,\x^{3}\,\xc^{2} +
1003413125\,\y\,\x^{3}\,\xc + 3385996250\,{\y}\,\x^{3} \\
 & & \mbox{} + 51211999525\,\x^{2}\,\xc^{2} +
16053417900\,\y\,\x^{2} + 4651759450\,\y\,{\x}^{2}\,\xc \\
 & & \mbox{} + 177479507850\,\x^{2}\,\xc +
140245780600\,\x^{2} + 212140488080\,\x \\
 & & \mbox{} + 7133355840\,\y\,\x\,\xc +
272934958520\,\x\,\xc + 25260582880\,\y\,{\x} \\
 & & \mbox{} + 79415365840\,\x\,\xc^{2} +
106967929856 + 13183919424\,\y + 41078949112\,\xc^{2}
 \\
 & & \mbox{} + 139995956352\,\xc + 3612843312\,\xc\,
\y, 427238747000\,\y\,\yc^{2} \\
 & & \mbox{} - 80593740760\,\y\,\yc - 961615825940\,
\yc^{2}\,\xc + 72041680708\,\yc\,\xc \\
 & & \mbox{} + 31587988349750\,\x^{2} + 9360849735875\,
\x^{2}\,\xc + 90434478667240\,\x \\
 & & \mbox{} + 26404464403460\,\x\,\xc + 549197297800
\,\x\,\yc + 64702426970096 \\
 & & \mbox{} + 18636073790528\,\xc + 1397883090496\,{\yc} - 2547657512880\,\yc^{2},  \\
 & & 31185310\,\y\,\yc\,\x + 52924196\,\y
\,\yc - 6814654\,\yc\,\xc - 716849880\,\x
^{2} \\
 & & \mbox{} - 440552505\,\x^{2}\,\xc - 2156806048\,
\x - 1352366658\,\x\,\xc - 45578530\,\x\,
\yc \\
 & & \mbox{} - 1620399064 - 1038603700\,\xc - 90980056\,
\yc, 505750\,\x^{2} + 149875\,\x^{2}\,\xc
 \\
 & & \mbox{} + 1539520\,\x + 440460\,\x\,\xc +
116450\,\yc\,\x\,\xc + 232900\,\x\,{\yc} + 1160352 \\
 & & \mbox{} + 317976\,\xc + 372640\,\yc + 186320\,
\yc\,\xc]\,.
\end{eqnarray*}
\normalsize The only sets where the solutions $\x=const.$,
$\y=const.$ are not obvious are $G_1$  and $G_4$. For $G_1$, if we
substitute $\x=-8/5$ into $N_1$ we get $195\y\yc+702/5=0$, which
is incompatible with the first equation of this set.

Let us consider now the  fourth and fifth equations in set $G_4$
given, respectively, by
\begin{eqnarray}
\la{60} \nonumber & & \mbox{} 31185310\,{\yc}\,{\y}\,{\x} +
52924196\,{\yc}\,{\y} - 6814654\,{\yc}\,{\xc} - 716849880\,{\x}^{2} \\
\nonumber
 & & \mbox{} - 440552505\,{\x}^{2}
\,{\xc} - 1352366658\,{\x}\,{\xc} - 2156806048\,{\x} - 45578530\,{\x}\,{\yc} \\
 & & \mbox{}  - 1038603700\,{\xc} - 1620399064 - 90980056\,{\yc}=0
\end{eqnarray}
and
\begin{eqnarray}
\la{61} \nonumber & & \mbox{}
 505750\,{\x}^{2} + 149875\,{\x}^{2
}\,{\xc} + 440460\,{\x}\,{\xc} + 116450\,{\yc}\,{\x}\,{\xc} + 1539520\,{\x}\\
 & & \mbox{}  + 232900\,{\x}\,{\yc}  317976\,{\xc} + 186320\,{\yc}\,{\xc}
 + 1160352 + 372640\,{\yc}=0\,.
\end{eqnarray}
By applying {\tt grobner} to (\ref{60}), (\ref{61}) and their
complex conjugates (in this case, two new equations), we obtain a
system of polynomials for which all solutions have $x_1$ and $x_2$
constant.

Let us consider now the special cases in which the denominators in
the previous
 expressions for $\Phi_{11}$ are zero.
The denominators of  (\ref{54}), (\ref{53}), and (\ref{55}) are
given, respectively, by
\begin{eqnarray}
\la{d1} \nonumber \lefteqn{{d_1} := 10868\,\yc^{2} + 772320\,\yc +
1020276\,\x\,\yc + 4085\,\yc^{2}\,\x}  \\
 & & \mbox{} -1158240\,\x^{2}-2048352\,\x - 214700\,\x^{3}
+335985\,\yc\,\x^{2} - 1191680\,,
\end{eqnarray}
\be \la{d2} { d_2} := 148960 + 42940\,\x^{2} + 10412\,\yc^{2}
 - 18564\,\yc - 13047\,\x\,\yc + 162944\,{\x}\,,
\ee
\begin{eqnarray}
\la{d3} \nonumber \lefteqn{{ d_3} := (\,148960 + 42940\,\x^{2} +
10412\, \yc^{2} - 18564\,\yc - 13047\,\x\,\yc + 162944\,\x\,)} \\
\nonumber
 & & \mbox{} (- 24320\,\yc^{2} + 7684576\,\yc +
9460852\,\x\,\yc + 2915745\,\yc\,\x^{2}
 - 194465152\,\x \\
 & & \mbox{} - 123050612\,\x^{2} - 102596352 - 25995895\,
\x^{3} +133000\,\yc^{2}\,\x)\,.
\end{eqnarray}
Applying $\ode$ to (\ref{d1}) we obtain
\begin{eqnarray}
\la{yd1} \nonumber
 & & \mbox{} 63729588\,\yc\,\x^{2} - 670120
\,\x^{2}\,\yc^{2} + 12371205\,\x^{3}\,\yc
 - 617652\,\yc^{3} + 18435168\,\x^{2} \\
\nonumber
 & & \mbox{} + 108263616\,\x\,\yc + 10424160\,{\x}^{3}
+ 10725120\,\x - 1048608\,\yc^{2} + 60746496\,
\yc \\
 & & \mbox{} - 383325\,\yc^{3}\,\x - 1749232\,{\yc
}^{2}\,\x + 1932300\,\x^{4}\,=0\,.\mbox{\hspace{177pt}}
\end{eqnarray}
Applying $\ode$ again on (\ref{yd1}), gives
\begin{eqnarray}
\la{yyd1} \nonumber
 & & \mbox{}- 71849032\,\yc^{3}\,\x -
2917197592\,\x^{2}\,\yc^{2} + 2126429184\,\yc\,
\x^{2} \\
\nonumber
 & & \mbox{} + 3683108352\,\x\,\yc + 357185400\,\x^{3}\,\yc
- 128701440\,\x^{2} - 125089920\,{\x}^{4} \\
\nonumber
 & & \mbox{} - 67408896\,\yc^{3} - 18590290\,\x^{2}\,
\yc^{3} + 6696360\,\yc^{4} + 4179810\,\x\,\yc^{4} \\
 & & \mbox{} - 221222016\,\x^{3} - 2418547200\,\yc^{2
} + 2059223040\,\yc - 22411650\,\x^{4}\,\yc \\
 & & \mbox{} - 614629870\,\x^{3}\,\yc^{2} - 23187600
\,\x^{5} - 4605932928\,\yc^{2}\,\x \,=0\,.
\end{eqnarray}
Using {\tt grobner} package on (\ref{d1}), (\ref{yd1}) and
(\ref{yyd1}) gives the empty set solution. Applying $\ode$ to
(\ref{d2}), gives
\begin{eqnarray}
\la{yd2} \nonumber
 & & \mbox{} - 128832\,\yc^{2} - 101344\,{\yc
}^{2}\,\x - 2591852\,\yc\,\x^{2} - 8248048\,
\x\,\yc - 257640\,\x^{3} \\
 & & \mbox{} - 6550592\,\yc - 977664\,\x^{2} - 20824
\,\yc^{3} - 893760\,\x\,=0\,.\mbox{\hspace{131pt}}
\end{eqnarray}
Applying $\ode$ again on (\ref{yd2}), gives
\begin{eqnarray}
\la{yyd2} \nonumber
 & & \mbox{} 62472\,\yc^{4} - 60094064\,\yc
\,\x^{2} + 35714228\,\x^{2}\,\yc^{2} + 2832764
\,\x^{3}\,\yc \\
\nonumber
 & & \mbox{} - 2838720\,\yc^{3} + 8043840\,\x^{2} -
210698752\,\x\,\yc + 8798976\,\x^{3} + 86775744
\,\yc^{2} \\
 & & \mbox{} - 171601920\,\yc - 1690904\,\yc^{3}\,
\x + 111204048\,\yc^{2}\,\x + 2318760\,\x ^{4}\,=0\,.
\end{eqnarray}
Using {\tt grobner} package on (\ref{d2}), (\ref{yd2}) and
(\ref{yyd2}) we obtain the empty set solution. We observe now that
one of the factors in $d_3$ is  $d_2$. Thus, if $d_3=0$, we need
to consider just the expression
\begin{eqnarray}
\la{73b} \nonumber
 & & \mbox{} 133000\,\yc^{2}\,\x - 24320\,
\yc^{2} + 7684576\,\yc + 9460852\,\x\,\yc
 + 2915745\,\yc\,\x^{2}\\
 & & \mbox{} - 194465152\,\x - 123050612\,\x^{2} -
102596352 - 25995895\,\x^{3}=0\,.\mbox{\hspace{90pt}}
\end{eqnarray}
Applying $\ode$ twice on this equation, results in
\begin{eqnarray}
\la{69b} \nonumber
 & & \mbox{} 1923742456\,\yc\,\x^{2} -
54838615\,\x^{2}\,\yc^{2} + 387128860\,\x^{3}\,
\yc + 28673280\,\yc^{3} \\
\nonumber
 & & \mbox{} + 1750186368\,\x^{2} + 3181762656\,\x\,
\yc + 1107455508\,\x^{3} + 923367168\,\x \\
\nonumber
 & & \mbox{} - 143083296\,\yc^{2} + 1751900928\,\yc
 + 17772600\,\yc^{3}\,\x - 175990444\,\yc^{2}\,
\x \\
 & & \mbox{} + 233963055\,\x^{4}\,=0\mbox{\hspace{305pt}}
\end{eqnarray}
and
\begin{eqnarray}
\la{70b} \nonumber
 & & \mbox{} 3071183072\,\yc^{3}\,\x -
65381238064\,\x^{2}\,\yc^{2} + 363466389312\,{\yc}\,\x^{2} \\
\nonumber
 & & \mbox{} + 414477092352\,\x\,\yc + 141775882688\,
\x^{3}\,\yc - 11080406016\,\x^{2} \\
\nonumber
 & & \mbox{} - 13289466096\,\x^{4} + 2839158528\,\yc
^{3} + 811332320\,\x^{2}\,\yc^{3} - 315187200\,{\y}^{4} \\
\nonumber
 & & \mbox{} - 196695600\,\x\,\yc^{4} - 21002236416\,
\x^{3} - 58327724544\,\yc^{2} + 177286496256\,{\yc} \\
\nonumber
 & & \mbox{} + 20770389380\,\x^{4}\,\yc - 13320040240
\,\x^{3}\,\yc^{2} - 2807556660\,\x^{5} \\
 & & \mbox{} - 106950649152\,\yc^{2}\,\x\,=0\,.
\end{eqnarray}\,
Using {\tt grobner} package to (\ref{69b}) and (\ref{70b}) we
obtain again the empty solution set.

We need now to study the case in which $x_1$ and $x_2$ are
constants. From $\ode x_1 = \ode(\al/\pi)=0$ and $\de x_2 =
\de(\beta/\opi)=0$ we get, respectively,
 \be
  \la{75a}
31\,\x+10\,{\x}^{2}+2=0 \ee and \be \la{75b}
63\,{\y}+40\,\y\,\xc+\xc\,=0\,.
 \ee
 The above equations  have two solutions,
given by $x_1=-3/2$, $x_2=1/2$ and $x_1=x_2=-8/5$. The first one
satisfies $5\x+\yc-8=0$, which will be considered next. The second
case is impossible, since these values don't make $N_1$ equal to
zero.

Let us now consider the case:
 \be
  \la{81}
   \beta-5\oal-8\opi=0\,,
\ee
 or, using variables $\x$ and $\y$, and dividing by $\pi$,
  \be
\la{82}
 8-\y-5\x=0\,.
 \ee
 Applying $\de$ to this equation, using
(\ref{32}),(\ref{51}) and (\ref{52}), we get
 \be
 \la{83} 34 \y \xc
-\y^2+56\y+15\xc^2+24\xc=0\,. \ee
The only solution for both
equations is given by
 \be
 \la{86a}
 x_1=-3/2\,,\qquad x_2=1/2\,.
\ee
 Since the numerator on the right side of (\ref{48}) must be
zero, we obtain, solving for $\Phi_{11}$,
\be
 \la{86}
\Phi_{11}=-\frac{86}{141}\pi \opi\,.
\ee
 Applying $\delta$ on this
equation, we obtain
\be
 \la{87}
\delta \pi =-\pi \opi\,.
\ee
 Equations (\ref{37b}) and  (\ref{86a}) give
  \be
   \la{87b} \ode \pi =
\pi^2\,.
 \ee
 Computation of  the commutator $[\ode, \delta]\pi $,
using (\ref{87}) and (\ref{87b}) gives $\pi=0$.

Finally, we consider the case $\Phi_{11}=0$. Here both
denominators of (\ref{54}), and of  (\ref{55}) must be equal to
zero. Applying Buchberger-Gr\"obner algorithm to them and their
complex conjugates and using the variables defined in (\ref{x_1}),
we can verify that the only possible solutions again fall in cases
we have studied before, i.e., either $x_1$ or $x_2$ are zero, or
both $x_1$ and $x_2$ are constants.

\section{\bf Conclusions}
Theorem 1 was  proved for the case of Maxwell equations, i.e.,
\emph{there
 are no Petrov type III space-times on which
 Maxwell's equations satisfy Huygens' principle.}
For the neutrino case, $k_1=8$, $k_2=13$, the proof is similar
\cite{sasse97}.  The use of the package {\tt NPpsinor} in Maple
was essential for the conversion of the Alvarez-W\"{u}nsch
five-index necessary condition from the tensorial to dyad form.

The polynomial system obtained from the integrability conditions
was simplified  using   Maple's package {\tt grobner}. Since a
direct application of the algorithm is apparently impossible due
to the large size of the polynomial system, a ``divide and
conquer'' approach was applied with success, showing that the
necessary conditions III and V for the validity of Huygens'
principle cannot be simultaneously satisfied for Maxwell's
equations in Petrov type III space-times.
\\  \\
\noindent
 \Large

{\bf References}
  \normalsize

\begin{thebibliography}{References}
\bibitem{alv91a} {\sc M. Alvarez}, {\sl Zum huygensschen Prinzip bei einigen Klassen
spinorieller Feldgleichungen in gekr\"{u}mmten
Raum-Zeit-Manifaltigkeiten}, Diss. A. P\"{a}d. Hochshule
Erfurt/M\"{u}hlhausen, 1991.

\bibitem{alv91b} {\sc M. Alvarez} and {\sc V. W\"{u}nsch}, Zur G\"{u}ltigkeit des
huygensschen Prinzips bei der Weyl-Gleichung und den homogenen
Maxwellschen Gleichungen f\"{u}r Metriken vom Petrov-typ N, {\sl
Wiss. Zeitschr. P\"{a}d. Hochsch. Erfurt/M\"{u}hlhausen,
Math.-Naturwiss}. {\bf 27}, 1991, pp. 77-91.



\bibitem{car84} {\sc J. Carminati} and  {\sc R. G. McLenaghan}, Determination of
all Petrov type N space-times on which the conformally invariant
scalar wave equation  satisfies Huygens' principle, {\sl Phys.
Lett.}{105A}, 1984, pp. 351-354.

\bibitem{car86} {\sc J. Carminati} and  {\sc R. G. McLenaghan}, An explicit
determination of the  Petrov type N space-times on which the
conformally invariant scalar wave equation  satisfies Huygens'
principle, {\sl Ann. Inst. Henri Poincar\'{e}, Phys.
Th\'{e}or.}{\bf 44}, 1986, pp. 115-153.

\bibitem{car87} {\sc J. Carminati} and  {\sc R. G. McLenaghan}, An explicit
determination of space-times on which the conformally invariant
scalar wave equation  satisfies Huygens' principle. Part II:
Petrov type D space-times, {\sl Ann. Inst. Henri Poincar\'{e},
Phys. Th\'{e}or.} {\bf 47}, 1987, pp. 337-354.

\bibitem{car88} {\sc J. Carminati} and  {\sc R. G. McLenaghan}, An explicit
determination of space-times on which the conformally invariant
scalar wave equation  satisfies Huygens' principle. Part III:
Petrov type III space-times, {\sl Ann. Inst. Henri Poincar\'{e},
Phys. Th\'{e}or.} {\bf 48}, 1988, pp. 77-96.


\bibitem{cza87} {\sc S. R. Czapor} and {\sc R. G. McLenaghan}, NP: A Maple
package for performing calculations in the Newman-Penrose
formalism. {\sl Gen. Rel. Gravit.} {19}, 1987, pp. 623-635.

\bibitem{cza92} {\sc S. R. Czapor, R. G. McLenaghan} and {\sc J. Carminati},
The automatic conversion of spinor equations to dyad form in
MAPLE, {\sl Gen. Rel. Gravit.} {24}, 1992, pp. 911-928.


\bibitem{gun52} {\sc P. G\"{u}nther}, Zur G\"{u}ltigkeit des huygensschen
Prinzips bei partiellen Differentialgleichungen von normalen
hyperbolischen Typus, {\sl S.-B. Sachs. Akad. Wiss. Leipzig
Math.-Natur K.}, {\bf 100}, 1952, pp. 1-43.


\bibitem{gun65b} {\sc P. G\"{u}nther}, Ein Beispiel einer nichttrivialen
huygesschen Differentialgleichungen mit vier unabh\"{a}ngigen
Variablen, {\sl Arch. Rational Mech. Anal.} {\bf 18}, 1965, pp.
103-106.

\bibitem{gun74} {\sc P. G\"{u}nther and V. W\"{u}nsch}, Maxwellsche
Gleichungen und huygensches Prinzip I. {\sl Math. Nachr}. {\bf
63}, 1974, pp. 97-121.

\bibitem{had23} {\sc J. Hadamard}, {\sl Lectures on Cauchy's problem in linear
differential equations}. Yale University Press, New Haven, 1923.


\bibitem{kun68} {\sc H. P. K\"{u}nzle}, Maxwell fields satisfying Huygens'
principle, {\sl Proc. Cambridge Philos. Soc.}{\bf 64}, 1968, pp.
779-785.


\bibitem{mcl69} {\sc R. G. McLenaghan}, An explicit determination of the
empty space-times on which the wave equation satisfies Huygens'
principle,
 {\sl Proc. Cambridge Philos. Soc.}{\bf 65}. 1969, pp. 139-155.

\bibitem{mcl72} {\sc R. G. McLenaghan} and {\sc J. Leroy}, Complex recurrent
space-times, {\sl Proc. Roy. Soc. London} {A327}, 1972, pp.
229-249.

\bibitem{mcl74} {\sc R. G. McLenaghan}, On the validity of Huygens' principle
for second order partial differential equations with four
independent variables. Part I: Derivation of necessary conditios.
{\sl Ann. Inst. Henri Poincar\'{e}} {\bf A20}, 1974, pp.153-188.



\bibitem{mcl90} {\sc R. G. McLenaghan} and {\sc T. G. C. Williams},
An explicit determination of the Petrov type D space-times on
which Weyl's neutrino equation and Maxwell's equations satisfiy
Huygens' principle, {\sl Ann. Inst. Henri Poincar\'{e}, Phys.
Th\'{e}or.}{\bf 53}, 1990, pp. 217-223.

\bibitem{sasse97}
{F. D. Sasse}, {\sl Huygens' Principle for Relativistic Wave
Equations in Petrov Type III Space-Times}, Ph.D. Thesis,
University of Waterloo, Waterloo, 1997.

\bibitem{wun70} {\sc V. W\"{u}nsch}, \"{U}ber selbstadjungierte
Huygenssche Differentialgleichungen mit vier unabh\"{a}ngigen
Variablen, {\sl Math. Nachr.}, {\bf 47}, 1970, pp. 131-154.


\bibitem{wun78} {\sc V. W\"{u}nsch}, Cauchy-Problem und huygenssches Prinzip
bei einigen Klassen spinorieller Feldleichungen I. {\sl Beitr. zur
Analysis} {\bf 12}, 1978, pp. 47-76.

\bibitem{wun79} {\sc V. W\"{u}nsch}, Cauchy-Problem und huygenssches Prinzip
bei einigen Klassen spinorieller Feldleichungen II. {\sl Beitr.
zur Analisys} {\bf 13}, 1979, pp. 147-177.

\bibitem{wun89} {\sc V. W\"{u}nsch}, Huygens' principle on Petrov type-D
space-times {\sl Ann. Phys.} {\bf 46}, 1989, pp. 593-597.

\end{thebibliography}
\end{document}